%% file: main.tex
\pdfoutput=1

\documentclass[journal]{IEEEtran}

\usepackage{style}

\justifying

\input{Sections/00.metadata.tex}

\begin{document}
\maketitle

\begin{abstract}
\input{Sections/01.abstract.tex}
\end{abstract}


\section{Introduction}
\input{Sections/03.introduction.tex}

\section{Methodology}
\input{Sections/04.methodology.tex}

\section{Risk Evaluation}
\input{Sections/05.risk.tex}

\section{Results and Discussion}
\input{Sections/06.results.tex}

\section{Conclusion}
\input{Sections/07.conclusion.tex}

\bibliographystyle{IEEEtran}
\bibliography{references}

\end{document}

%% file: Sections/00.metadata.tex
\title{\textbf{Multilevel Verification on a Single Digital Decentralized Distributed (DDD) Ledger}}

\author{%
    Ayush Thada, Aanchal Kandpal, and Dipanwita Sinha Mukherjee%
    \thanks{Ayush Thada is with Wells Fargo India \& Philippines, Bengaluru, India (e-mail: ayush.thada@wellsfargo.com).}%
    \thanks{Aanchal Kandpal is with Wells Fargo India \& Philippines, Bengaluru, India (e-mail: aanchal.kandpal@wellsfargo.com).}%
    \thanks{Dipanwita Sinha Mukherjee is with Wells Fargo India \& Philippines, Bengaluru, India (e-mail: dipanwita.sinhamukherjee@wellsfargo.com).}%
}

\date{}

%% file: Sections/01.abstract.tex
This paper proposes a method for incorporating decentralized distributed digital (DDD) ledgers, such as blockchain, with multilevel verification. Traditional DDD ledgers like blockchain typically rely on a single verification level, limiting their utility in systems with hierarchical structures where multiple levels of verification are required. In systems where hierarchies naturally arise, integrating hierarchical verification offers a more robust solution. This involves both verifications within individual levels and across different levels, which introduces new challenges. For instance, each level must validate the work of the preceding level, ensuring consistency throughout the hierarchy. This paper addresses these challenges and outlines a framework for tracking the system's state at any given point and assessing its likelihood of failure.

%% file: Sections/02.keywords.tex
Blockchain, Decentralized Ledger, Cryptocurrency, Multilevel Verification, Distributed Systems, Decentralized Systems, Consensus Mechanism, Hierarchical Verification, Smart Contracts, Distributed Ledger Technology, Blockchain Scalability, Peer-to-Peer Networks, Digital Trust

%% file: Sections/03.introduction.tex
Ledgers have served as foundational record-keeping systems for centuries, but with the rise of decentralized technologies, they need to evolve to ensure secure and efficient data verification. A major breakthrough in this area came with the introduction of blockchain by Satoshi Nakamoto \cite{1}, where cryptographic principles such as hashing and proof-of-work enabled decentralized, tamper-resistant transactions. This innovation revolutionized finance by introducing Bitcoin, a new form of digital currency, and laid the groundwork for decentralized applications.

Building on Nakamoto’s original concept, further research has explored blockchain’s trade-offs and challenges. Monrat et al. \cite{2} provide a detailed survey on blockchain applications, highlighting scalability, privacy, and regulatory challenges. Zheng et al. \cite{3} and Wüst and Gervais \cite{4} extend this by categorizing blockchain use cases and analyzing consensus mechanisms critical for maintaining security in decentralized environments. Zhou et al. \cite{5} specifically address scalability issues, offering solutions such as off-chain scaling techniques to enhance blockchain’s efficiency.

Ethereum \cite{6}, introduced by Vitalik Buterin, expanded on blockchain technology by integrating smart contracts, allowing for automated, multi-step processes on the blockchain itself. However, the concept of smart contracts was initially developed by Nick Szabo \cite{7}, who envisioned self-executing contracts to enhance the security and efficiency of digital agreements. Szabo’s work laid the foundation for Ethereum’s smart contract platform, enabling decentralized applications (dApps) to flourish \cite{8}. Christidis and Devetsikiotis \cite{9} explored the potential of combining blockchain with the Internet of Things (IoT), while Cong and He \cite{10} analyzed blockchain’s disruptive potential in financial markets, focusing on the implications of decentralized contracts on transaction costs. Wang et al. \cite{11} provide an in-depth exploration of blockchain-enabled smart contract architecture and applications, while Alharby and Moorsel \cite{12} focus on systematically mapping the research landscape, identifying gaps, and categorizing the challenges and solutions in blockchain-based smart contracts.

Blockchain’s impact is most visible in the financial sector, where its ability to streamline processes and reduce intermediary involvement has driven widespread adoption. Varma \cite{13} reviews blockchain’s role in transforming financial systems, from peer-to-peer payments to decentralized finance (DeFi). Similarly, Lewis et al. \cite{14} and Zhang et al. \cite{15} explore how blockchain is reshaping financial market innovation by enhancing transparency and reducing transaction times. Schär \cite{16} discusses the transformation of financial markets through DeFi enabled by blockchain and smart contracts, while Vysya and Kumar \cite{17} review blockchain adoption in financial services, and Patel et al. \cite{18} provide a bibliometric analysis of blockchain research trends in banking and finance. Beyond finance, blockchain’s potential extends to healthcare, where it offers decentralized solutions for securely managing clinical trials, electronic health records, and regulatory compliance, as discussed by Hasselgren et al. \cite{19} and Mettler \cite{20}.

Despite these advances, existing work predominantly focuses on single-level verification processes in decentralized ledgers. Even studies that address multilevel authentication and blockchain applications, such as those by Balakrishnan et al. \cite{21} and Mbarek et al. Hölbl et al. \cite{22} systematically review the use of blockchain in healthcare, emphasizing data security and privacy, while Ben Fekih and Lahami \cite{23} provide a comprehensive study on blockchain applications in healthcare, focusing on electronic health records, supply chains, and clinical trials. Mbarek et al. \cite{24} propose a multilevel system via the use of agents, but blockchain-based verification is still limited to a single level within a hierarchy. Our previous work \cite{25} introduced a two-level verification system, where two authorities were responsible for authentication. However, this approach was confined to a specific application and lacked generality.

In this paper, we address this gap by presenting a generalized framework for hierarchical verification in decentralized ledgers. Unlike previous studies, our framework supports multilevel verification across hierarchical structures, making it applicable across diverse domains, including finance, healthcare, and IoT.

%% file: Sections/04.methodology.tex
\subsection*{Background Knowledge}

Before presenting our solution, we provide a brief overview of key terminologies used throughout the paper.\\

\begin{itemize}
    \item \textbf{Proof of Work (PoW)} \\
    Proof of Work is a consensus algorithm widely used in cryptocurrencies such as Bitcoin and Ethereum to validate transactions and add blocks to a decentralized ledger. The concept was first proposed by Dwork and Naor \cite{26}, and the term "proof of work" was later popularized by Jakobsson and Juels \cite{27}. PoW involves solving complex mathematical puzzles that require significant computational resources but are straightforward to verify. These puzzles ensure that all nodes in the network reach a consensus on the legitimacy of a new block. 

    The most commonly used form of PoW involves hash-based puzzles, where the goal is to find a partial hash inversion \cite{29} \cite{30}. Approaches such as Merkle-tree based puzzles \cite{27}, and Diffie-Hellman-based challenges \cite{28} have also been explored, but hash sequences \cite{31} remain predominant in modern cryptocurrency systems. While PoW secures decentralized networks, it has several drawbacks. The computational complexity of solving PoW puzzles leads to high energy consumption, contributing to significant environmental impacts due to increased carbon emissions. Moreover, PoW’s reliance on distributed verification increases latency, particularly in systems requiring multilevel verification, as in hierarchical blockchain systems. In addition, PoW is vulnerable to the "51\% attack" scenario, where an attacker with control over more than half of the network’s nodes can manipulate the ledger’s data \cite{1}.\\

    \item \textbf{Systemic Hierarchies} \\
    Systemic hierarchies refer to structured arrangements of entities—whether objects, roles, or individuals—within a system, where decision-making authority is distributed across different levels. In decentralized systems, multiple entities may be responsible for making decisions regarding the same objective. Conflicts can arise when decisions differ, necessitating higher-level authorities to resolve these discrepancies. As described by Balakrishnan et al. \cite{21}, multilevel verification systems leverage these hierarchies to manage complex decision-making processes.

    In our work, we assume a hierarchical structure where the number of decision-making entities decreases at each higher level. By the top of the hierarchy, decision conflicts are resolved entirely. This hierarchical approach helps reduce bias in decision-making by distributing responsibility across multiple levels while ensuring consistency at the top \cite{24}. Additionally, systemic hierarchies allow for more granular verification processes at each level, improving the overall reliability and security of the ledger.
\end{itemize}

\subsection*{Proposed Framework}

\begin{figure}[htp]
    \centering
    \includegraphics[width=\columnwidth]{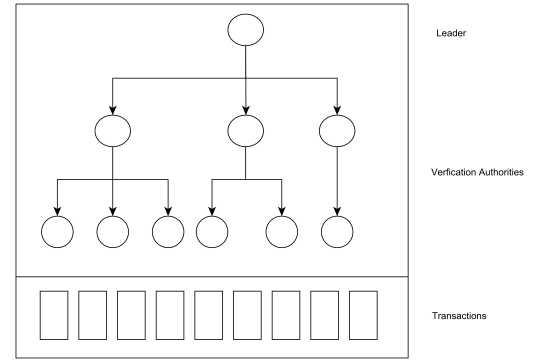}
    \caption{Hierarchy of Authorities}
    \label{fig:tree}
\end{figure}

To use the DDD ledgers for multilevel verification, we need to define certain components that, once arranged and used in a deterministic manner, will help us achieve multilevel authentication over the existing DDD ledgers. The first component is the nodes. Node here refers to the use case information like transactions and other verification metadata information stored in the ledgers. The second crucial component of this framework is hierarchy. For a given DDD ledger network, the structure of the hierarchy will be fixed. Each node in the hierarchy represents a verification agency. For a given DDD ledger network, the verification agencies will be chosen randomly. In the paper, the highest authority in the hierarchy is referred to as a leader, and the remaining authorities in all other levels as authorities. The nomenclature of the entire hierarchy is depicted in \hyperref[fig:tree]{Figure 1}.

\begin{figure}[htp]
    \centering
    \includegraphics[width=\columnwidth]{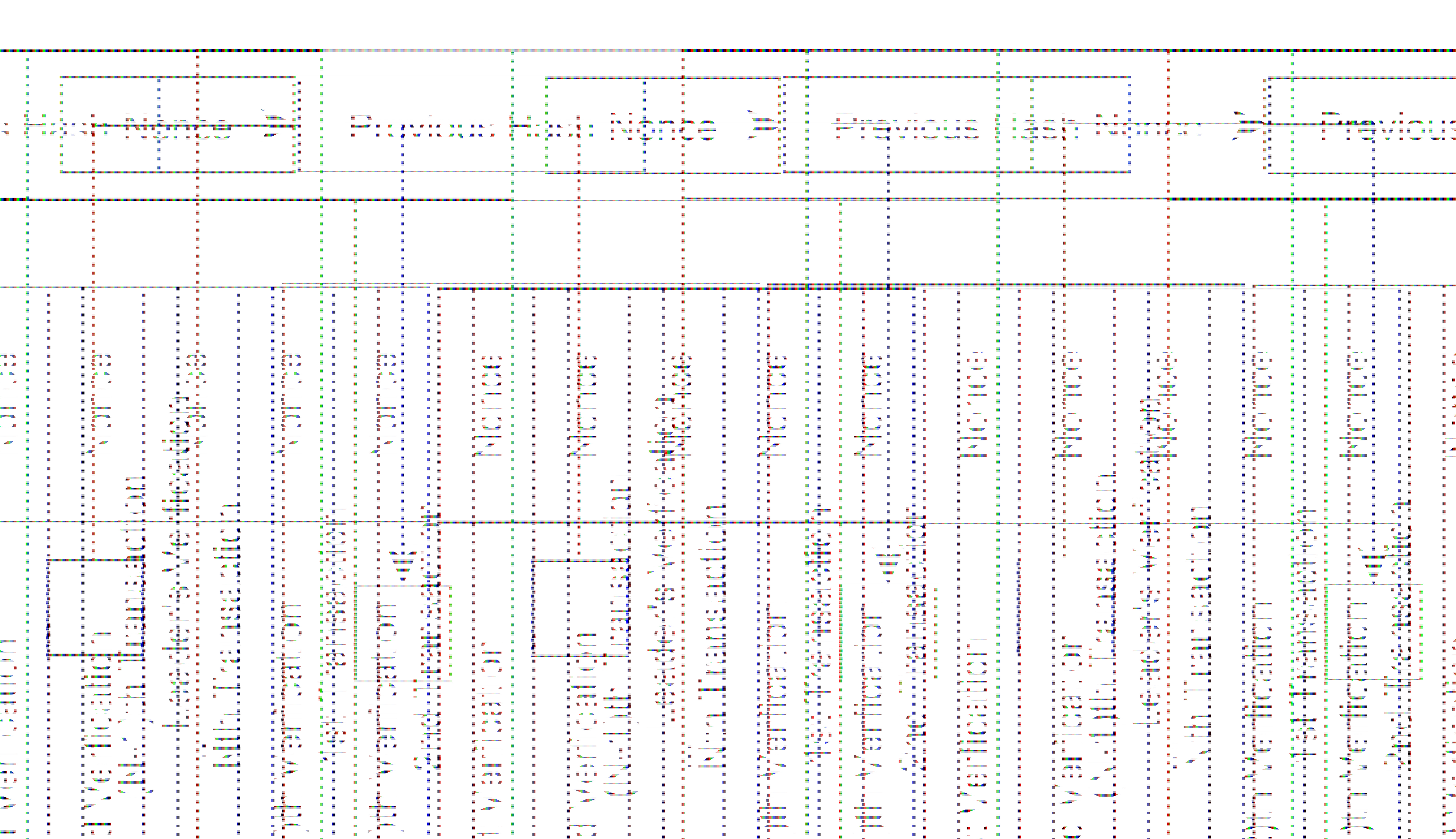}
    \caption{Single Block of Proposed DDD Ledger}
    \label{fig:block}
\end{figure}

For each level, there must be at least one authority. Hence, there will be at least one node. Here, each node will store different information and will represent the different states of the system. For a given level of the hierarchy, if we want N verification, we require N+1 nodes for that level. The first node stores the actual transaction information, and the next N-1 nodes are created to store the verification details by an authority. The last node stores the information of the verification by the leader. We can classify nodes into three categories: transaction node, which is submitted by a client; transaction metadata node, which is submitted by an authority; and block metadata node, which a leader submits. The difference here from the blockchain is that in the blockchain, a single block contains transactions, i.e., transaction nodes. However, in our framework, the block will contain information on transactions, information related to verification by the authorities, and information regarding the verification by the leader. For a given system, we can set the maximum number of transactions that a block can hold as well as the waiting time up to which transactions can be added so that in case the transactions are less than the max limit, the system will not be in the idle state forever, and the submitted transaction will be verified in time. There could be more than one transaction node in a block, but only one metadata node for each authority verifies the transaction/transactions. Within the same level in the hierarchy, the ordering of the metadata nodes doesn't matter, but across the different levels in the hierarchy, there is a defined order of the metadata nodes. Verification starts from the bottom of the hierarchy and will move to the top. Hence, the transaction metadata nodes will be added in that order. Here, hierarchy only dictates the structure in which verification will be carried out, but it never specifies the agents of verification. Also, the method of verification will vary from level to level to ensure that the transaction is valid from different perspectives. \hyperref[fig:block]{Figure 2} represents a complete filled block of the blockchain.

\begin{figure}[htp]
    \centering
    \includegraphics[width=\columnwidth]{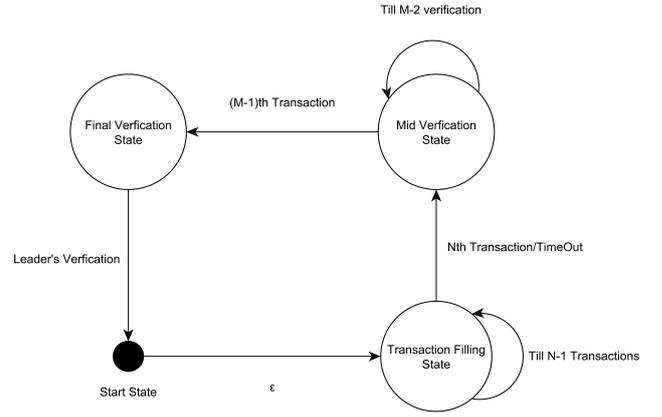}
    \caption{State Diagram of Proposed DDD Ledger}
    \label{fig:state}
\end{figure}

\hyperref[fig:state]{Figure 3} shows the three states in which our system can exist. The first stage is the transaction block-filling stage. In this stage, only transaction nodes will be added to the block up to a maximum limit for the block or the waiting time before transactions are sent for verification. The second stage is the verification stage. In this stage, verification authorities will verify the transactions, and the transaction metadata node will be added to contain their approval/disapproval of each transaction. Here, nonce will be added as per standard blockchain procedure. Along with their approval, they must show proof of the work that will be added to their metadata node. Before authorities of higher level proceed with their verification, they must first verify the metadata nodes added by the immediate lower level. Here, the original transaction serves as the ground truth for verification by authorities. Hence, each node in the hierarchy will verify the transaction. The only benefit we get out of multiple verifications is that the original transaction will be verified several times with different verification methods, which ensures that the transaction is valid. This ensures a linear amount of verification at each level. The verification will be done in the reverse order of nodes added to the block. The third stage is the final verification stage. In this stage, the leader will verify all authorities who have added their metadata hierarchy. Once the leader verifies all nodes in a block, the block metadata node will be added. The leader also needs to submit the proof of work, which will be common to both the block and the last node added to the block. After this block can be added to the DDD ledger, the system will automatically revert to the transaction filling stage.
 
 If there is more than one leader in the highest hierarchy, we need to create an additional hierarchy, let's call it aka\textunderscore leader, which logically produces the verification status of leaders. Once the hierarchy is updated, the aka\textunderscore leader will act like the leader as we defined above, and the leaders will become authorities according to the previous hierarchy.

%% file: Sections/05.risk.tex
In this framework, which generalizes a blockchain-based system, the risks inherent to blockchain are also relevant. Given that this is a distributed system, an adversary controlling more computational resources than the honest authorities could attempt to subvert the verification process across multiple hierarchical levels. In such a case, an adversary could potentially validate blocks with incorrect information through dishonest authorities at multiple levels, ultimately adding an invalid block to the longest DDD ledger. To counteract this risk, we suggest implementing domain-specific incentives alongside transaction fees, as in the original blockchain model \cite{1}. However, since this framework extends beyond financial transactions, the structure and nature of incentives would vary based on the application.

The potential risk of an adversary outpacing the honest chain can be modeled as a Binomial Random Walk, with success and failure events corresponding to the honest and adversarial chains extending by one block, respectively. In the hierarchical framework, verification must occur at multiple levels, with honest authorities outnumbering dishonest ones at each level to ensure the integrity of the transaction. Similarly, for an adversary to succeed, a majority of dishonest authorities must exist at each hierarchical level.\\

\subsubsection*{Hierarchical Model for Risk Assessment}

We assume a hierarchical structure modeled as a perfect n-ary tree of height $l$, where each node has $n$ children, except the leaf nodes, which have none. The total number of authorities, dishonest authorities, and honest authorities at each hierarchical level are represented as follows for $n > 1$:

\begin{equation}
    t = \frac{n^l-1}{n-1}
\end{equation}
\begin{equation}
    d = 1 + \Bigg\lceil\frac{n+1}{2}\Bigg\rceil\frac{n^{l-1}-1}{n-1}
\end{equation}
\begin{equation}
    h = \Bigg\lceil\frac{n-1}{2}\Bigg\rceil\frac{n^{l-1}-1}{n-1}
\end{equation}

These equations represent the total number of authorities ($t$), dishonest authorities ($d$), and honest authorities ($h$) at each level of the hierarchy. For $n = 1$, the values simplify to $t = d = 1$ and $h = 0$. These calculations are crucial in understanding the distribution of verification power across the hierarchical structure.\\

\subsubsection*{Probability of Adversarial Success}

To quantify the risk of an adversary catching up with the honest chain, we introduce the following probabilities:

- $p$: Probability that an honest authority correctly verifies a transaction. 

- $q$: Probability that a dishonest authority incorrectly validates an invalid transaction.

- $\bar{p}$: Probability that a correctly validated block is added to the longest DDD ledger.

- $\bar{q}$: Probability that an incorrectly validated block is added to the longest DDD ledger.

- $\bar{q}_z$: Probability that an adversary catches up from $z$ blocks behind.\\

In the hierarchical model, authorities at different levels verify the same transaction independently, and the process can vary at each level, which adds additional complexity to the validation. For $n > 1$, the probabilities $\bar{p}$ and $\bar{q}$ are computed as:

\begin{equation}
    \bar{p} = \mathds{1}_{(2\lceil\frac{n}{2}\rceil = n)}p^h(1-q) + \mathds{1}_{(2\lceil\frac{n}{2}\rceil \ne n)}p^h(1-q)^2
\end{equation}
\begin{equation}
    \bar{q} = q^{d}
\end{equation}

Here, $\bar{p}$ is the probability that a correctly verified block is added to the longest DDD ledger, taking into account the hierarchical structure. The indicator function $\mathds{1}$ adjusts the equation based on whether $n$ is even or odd, as this affects the majority requirement at each level.\\

For $n = 1$, 
\begin{equation}
    \bar{p} = (1-q)
\end{equation}
\begin{equation}
    \bar{q} = q
\end{equation}
where $q$ is the probability that a dishonest authority incorrectly verifies a block. \\

Using $\bar{p}$ and $\bar{q}$, we can compute $\bar{q}_z$, the probability that an adversary catches up from $z$ blocks behind:

\begin{equation}
    \bar q_z = \mathds{1}_{\bar{p}\le\bar{q}} + \mathds{1}_{\bar{p}>\bar{q}}\left(\frac{\bar{q}}{\bar{p}}\right)^z
\end{equation}

In this formula, the indicator function $\mathds{1}_{\bar{p} > \bar{q}}$ determines whether the adversary is likely to catch up based on the relative probabilities of honest versus dishonest authorities at various hierarchical levels.\\

\subsubsection*{Waiting Time Estimation}

An important aspect of risk evaluation is estimating how long a recipient should wait to be confident that a transaction cannot be reversed by an adversary. In a typical attack scenario, an adversary begins working on a parallel chain after submitting an initial transaction, attempting to overtake the honest chain by creating a longer fraudulent chain.

We model the adversary's progress as a Poisson distribution with an expected value is given as,
\begin{equation}
    \lambda = z\bar{q}_z
\end{equation}
where $z$ is the number of blocks added to the honest chain after the adversary's transaction. The probability of event $e$, that the adversary catches up and replaces the honest chain is:

\begin{equation}
\begin{aligned}
P(e|\lambda) &= \sum\limits_{k=0}^{\infty}\frac{\lambda^k e^{-\lambda}}{k!} \left(\bar{q}_{z-k}\mathds{1}_{k\le z} + \mathds{1}_{k>z}\right) \\
             &= \sum\limits_{k=0}^{z}\frac{\lambda^k e^{-\lambda}}{k!}\bar{q}_{z-k} + \sum\limits_{k=z+1}^{\infty}\frac{\lambda^k e^{-\lambda}}{k!}
\end{aligned}
\end{equation}

Taking our common factors, the final equation can be written as
\begin{equation}
    P(e|\lambda) = 1 - \sum\limits_{k=0}^{z}\frac{\lambda^k e^{-\lambda}}{k!}(1 - \bar{q}_{z-k})
\end{equation}

This equation calculates the probability that the adversary will still be able to catch up and reverse a transaction after $z$ blocks have been added to the honest chain. The Poisson distribution here models the adversary's potential progress over time, assuming the honest chain grows at its expected rate. The hierarchical verification model adds complexity, as the adversary must overcome multiple levels of verification to successfully rewrite the ledger.

%% file: Sections/06.results.tex
\subsection*{One-way Partial Dependence Analysis}

\begin{figure}[H] 
    \centering
    \begin{subfigure}[t]{\columnwidth}
        \centering
        \includegraphics[width=0.8\columnwidth, height=4.5cm]{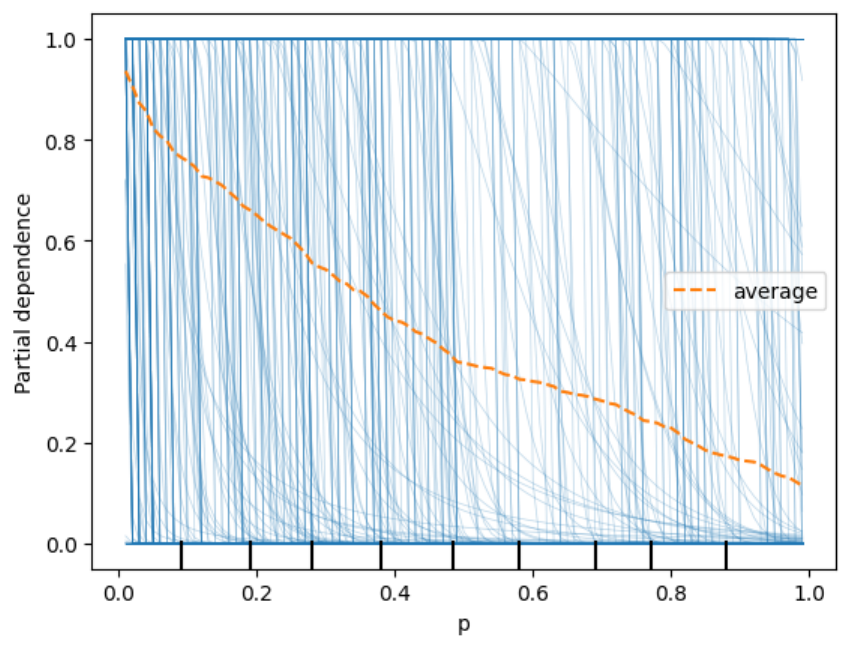}
        \caption{One-Way Partial Dependence Plot of $p$}
        \label{fig:ice1-1}
    \end{subfigure}
    
    \vspace{0.2cm}
    \begin{subfigure}[t]{\columnwidth}
        \centering
        \includegraphics[width=0.8\columnwidth, height=4.5cm]{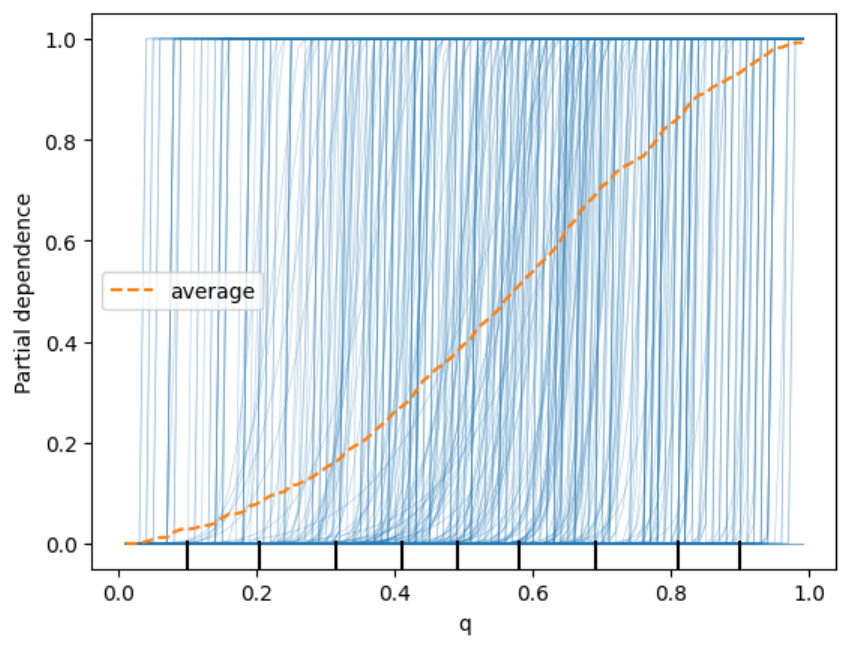}
        \caption{One-Way Partial Dependence Plot of $q$}
        \label{fig:ice1-2}
    \end{subfigure}
    
    \vspace{0.2cm}
    \begin{subfigure}[t]{\columnwidth}
        \centering
        \includegraphics[width=0.8\columnwidth, height=4.5cm]{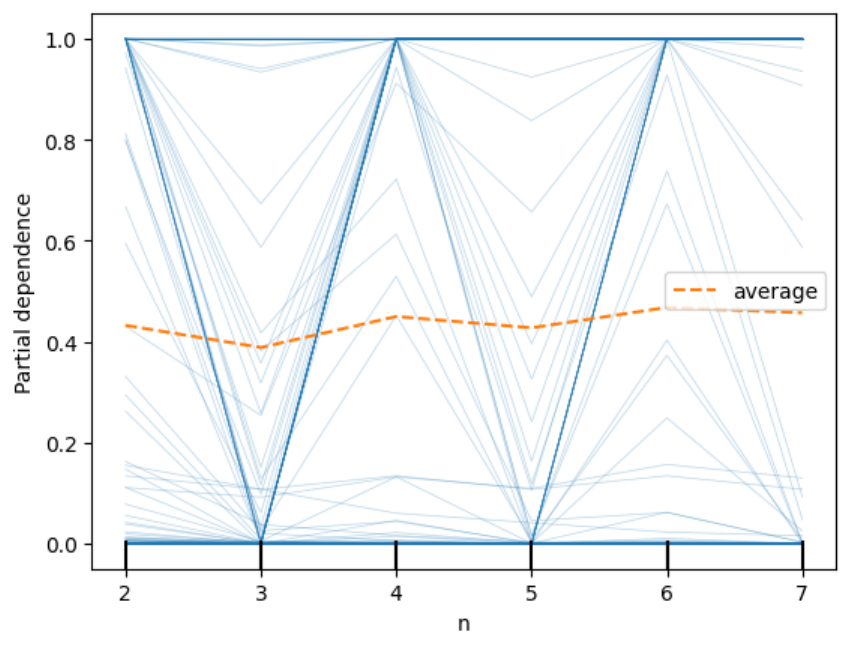}
        \caption{One-Way Partial Dependence Plot of $n$}
        \label{fig:ice1-3}
    \end{subfigure}
    
    \vspace{0.2cm}
    \begin{subfigure}[t]{\columnwidth}
        \centering
        \includegraphics[width=0.8\columnwidth, height=4.5cm]{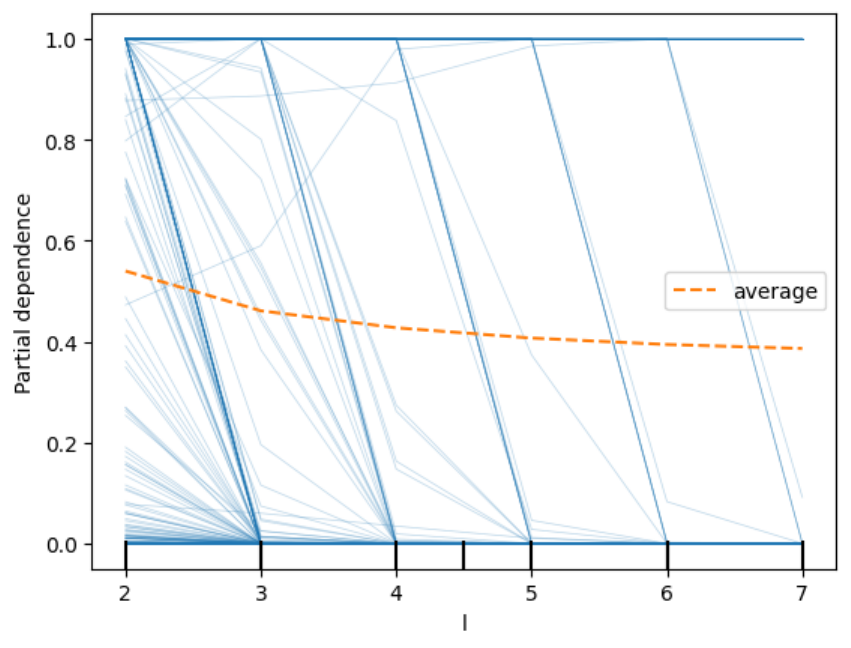}
        \caption{One-Way Partial Dependence Plot of $l$}
        \label{fig:ice1-4}
    \end{subfigure}
    
    \vspace{0.2cm}
    \begin{subfigure}[t]{\columnwidth}
        \centering
        \includegraphics[width=0.8\columnwidth, height=4.5cm]{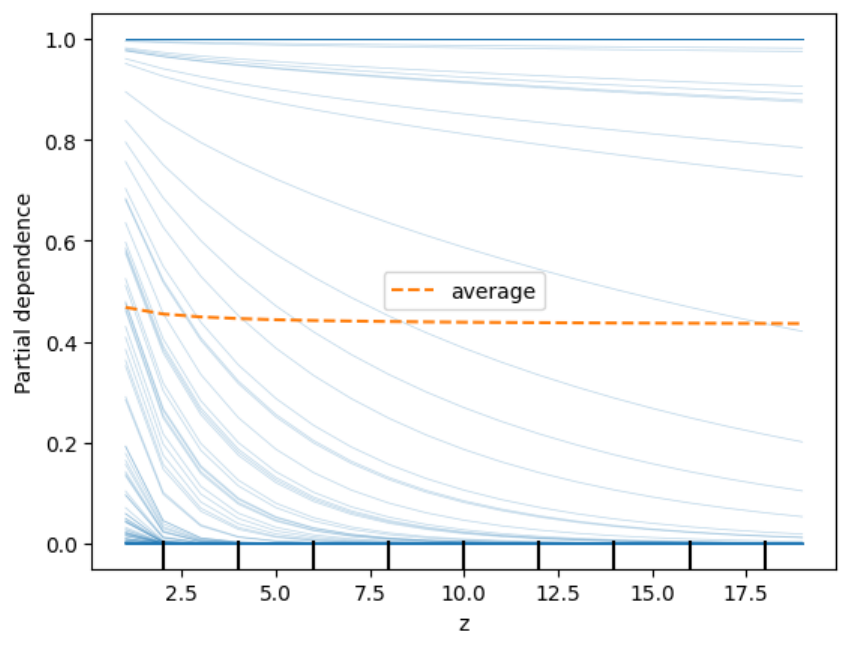}
        \caption{One-Way Partial Dependence Plot of $z$}
        \label{fig:ice1-5}
    \end{subfigure}
    \caption{One-way interaction Partial Dependence Plots}
    \label{fig:ice1}
\end{figure}

The behavior of the probability of the event $P(e|\lambda)$ is analyzed using one-way partial dependence plots (PDP) \cite{32} \cite{33} that show the average effect of each input variable on the final outcome. The combined PDP plot, which contains all the individual variable effects, is presented in \hyperref[fig:ice1]{Figure 4}.

From the PDP plots shown in \hyperref[fig:ice1]{Figure 4}, we can observe the following trends:

1. \textbf{Effect of $p$ (Probability of Correct Verification)}:
   - As the probability of correct verification $p$ increases, there is a steady decrease in $P(e|\lambda)$. This behavior is expected because as the honest authority becomes more likely to verify correctly, the adversary's chance of catching up diminishes. The diminishing probability indicates that the system becomes more secure with an increase in $p$.

2. \textbf{Effect of $q$ (Probability of Incorrect Verification)}:
   - As $q$, the probability of incorrect verification by a dishonest authority, increases, the value of $P(e|\lambda)$ increases. This suggests that with higher chances of incorrect verification, it becomes easier for an adversary to catch up, making the system more vulnerable.

3. \textbf{Effect of $n$ (Structure of the Hierarchy)}:
   - The plot shows a sawtooth-like pattern for the structure of the hierarchy, particularly in an $n$-ary tree, where `n` refers to the number of nodes at each level. Specifically, for even values of $n$, the probability decreases, while for odd values, it increases slightly. This behavior can be attributed to the fact that for a given even value of $n$, the number of verifications required stabilizes, but with odd values, the structure becomes less efficient, causing a temporary increase in the probability of adversary success.

4. \textbf{Effect of $l$ (Height of the Tree)}:
   - As the height of the tree, $l$, increases, we observe a decrease in $P(e|\lambda)$. This is intuitive since a taller hierarchy implies more verifications are required, making it progressively harder for the adversary to catch up.

5. \textbf{Effect of $z$ (Number of Honest Verifications)}:
   - For increasing values of $z$, the number of honest verifications, the plot shows a decreasing trend in $P(e|\lambda)$. This effect is consistent across different parameter choices, indicating that more honest verifications consistently lower the adversary’s probability of catching up.\\

Overall, the one-way partial dependence plots provide a clear understanding of how each individual variable affects the final probability of an adversary catching up. The results are consistent with the expected theoretical behavior, where increased honest participation and stricter verification mechanisms reduce the adversary's success probability.

\subsection*{Two-way Partial Dependence Analysis}

To better understand the interaction between variables and their combined effect on $P(e|\lambda)$, we present two-way partial dependence plots (PDPs). These plots display the average predicted value of $P(e|\lambda)$ for different combinations of input variables, allowing us to explore how pairs of variables interact. \\

The contour plots shown in \hyperref[fig:pdp2]{Figure 5} help illustrate how pairs of variables interact to influence the probability $P(e|\lambda)$. The color gradient represents the magnitude of $P(e|\lambda)$, with warmer colors (yellow) indicating higher values and cooler colors (blue/purple) indicating lower values. Each contour line represents a constant value of $P(e|\lambda)$. Key observations are as follows:\\

1. \textbf{Interaction between $p$ and $q$}:
   - The plot of $p$ versus $q$ shows diagonal contour lines, suggesting a strong interaction between these two variables. As both $p$ (probability of correct verification) and $q$ (probability of incorrect verification) increase, the probability $P(e|\lambda)$ also changes significantly. The diagonal lines indicate that both variables have a similar, compounding effect on $P(e|\lambda)$, meaning that changes in both $p$ and $q$ simultaneously have a significant impact on the adversary's ability to catch up.\\
   
2. \textbf{Interaction between $l$ and $n$}:
   - In the plot for $l$ (tree height) and $n$ (structure of the hierarchy), we observe a more complex pattern, with slanted contour lines in some regions and horizontal lines in others. This indicates that for certain ranges of $l$ and $n$, their combined effect on $P(e|\lambda)$ is nonlinear. The interaction becomes more pronounced at specific values of $n$, where an increase in $l$ can drastically reduce the adversary's success probability.\\
   
3. \textbf{Interaction between $n$ and $z$}:
   - The interaction between $n$ (number of nodes) and $z$ (number of honest verifications) shows a distinctive cyclic pattern. The contour lines are almost vertical for smaller values of $n$ and $z$, suggesting that $n$ has a dominant influence when $z$ is small. However, as $z$ increases, we observe cyclical shifts in the contour lines, indicating an alternating effect. For larger values of $z$, the level sets stabilize, meaning that the influence of $n$ on $P(e|\lambda)$ diminishes as the number of verifications increases.\\

4. \textbf{Interaction between $z$ and $q$}:
   - The plot of $z$ versus $q$ shows a combination of vertical and diagonal contours, implying a moderate interaction between these variables. As $q$ (incorrect verification) increases, the value of $P(e|\lambda)$ rises steeply for low values of $z$, indicating that the adversary’s success is highly sensitive to $q$ when fewer honest verifications are performed. However, for higher values of $z$, this sensitivity decreases.\\

5. \textbf{Interaction between $l$ and $z$}:
   - In the plot for $l$ versus $z$, the contours indicate that as both the height of the hierarchy ($l$) and the number of honest verifications ($z$) increase, the probability $P(e|\lambda)$ decreases. The contours are mostly vertical for lower values of $z$, indicating little interaction. However, at higher values of $z$, there is a stronger interaction effect, especially when $l$ increases, making it harder for the adversary to succeed.\\

The two-way partial dependence plots show significant interactions between key variables. The most prominent interaction is observed between $p$ and $q$, where changes in both probabilities jointly affect the outcome significantly. For other pairs, like $l-n$ and $n-z$, the interaction effects are more localized, with certain regions showing stronger interactions. These insights are crucial for understanding how different factors combine to affect the probability $P(e|\lambda)$ and how interventions targeting these factors can improve security.\\

\begin{figure}[H] 
    \centering
    \begin{subfigure}[t]{\columnwidth}
        \centering
        \includegraphics[width=0.8\columnwidth, height=4.5cm]{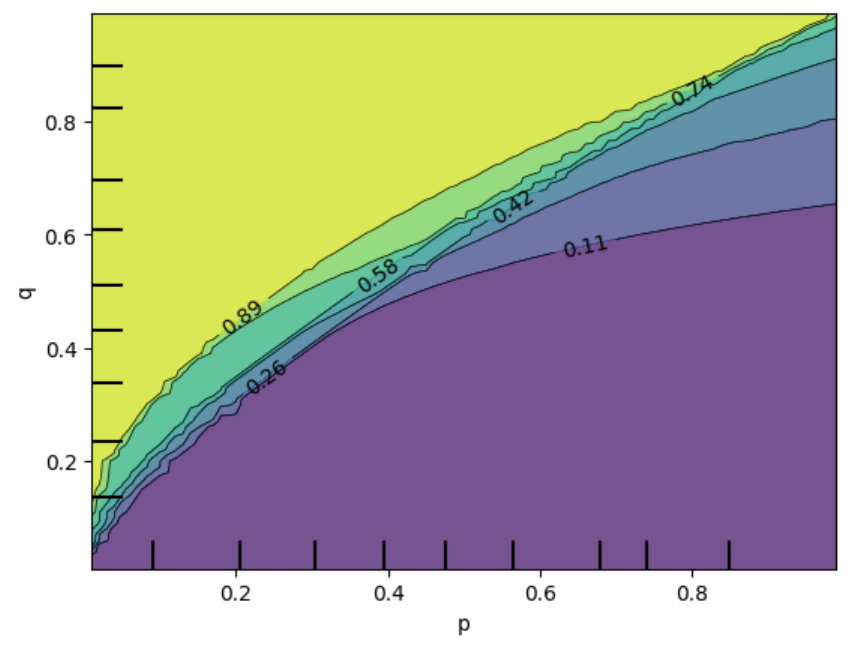}
        \caption{Two-Way Partial Dependence Plot between $p$ and $q$}
        \label{fig:pdp1-1}
    \end{subfigure}
    
    \vspace{0.2cm}
    \begin{subfigure}[t]{\columnwidth}
        \centering
        \includegraphics[width=0.8\columnwidth, height=4.5cm]{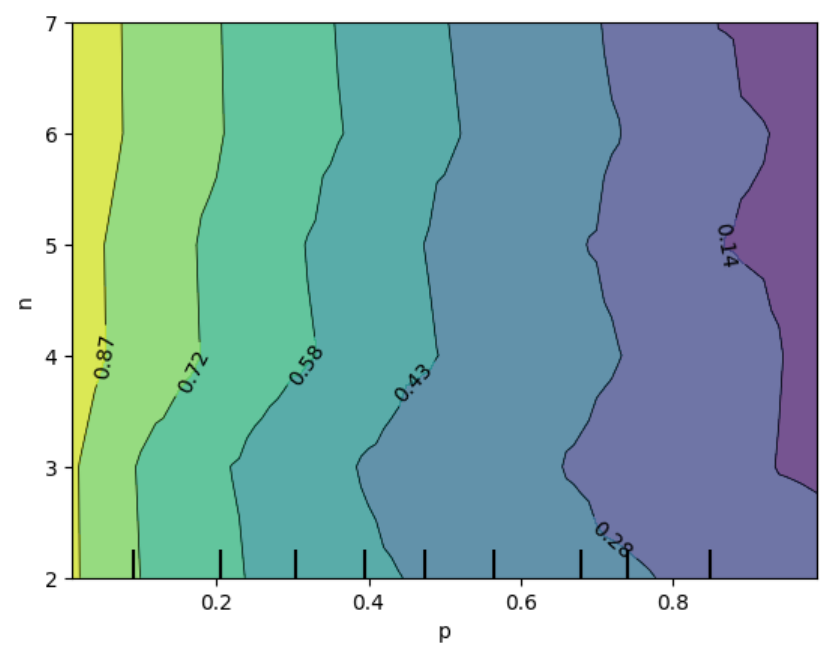}
        \caption{Two-Way Partial Dependence Plot between $p$ and $n$}
        \label{fig:pdp1-2}
    \end{subfigure}
    
    \vspace{0.2cm}
    \begin{subfigure}[t]{\columnwidth}
        \centering
        \includegraphics[width=0.8\columnwidth, height=4.5cm]{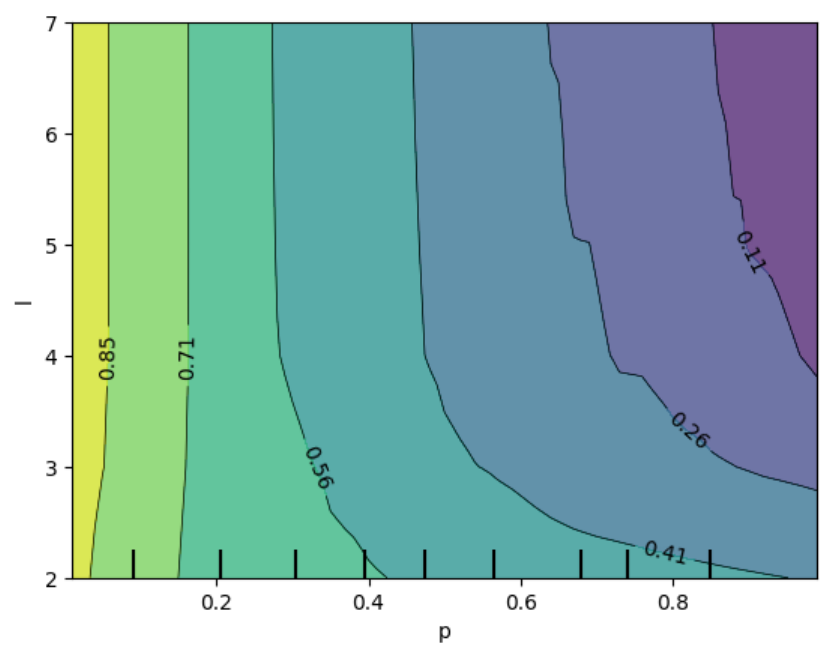}
        \caption{Two-Way Partial Dependence Plot between $p$ and $l$}
        \label{fig:pdp1-3}
    \end{subfigure}
    
    \vspace{0.2cm}
    \begin{subfigure}[t]{\columnwidth}
        \centering
        \includegraphics[width=0.8\columnwidth, height=4.5cm]{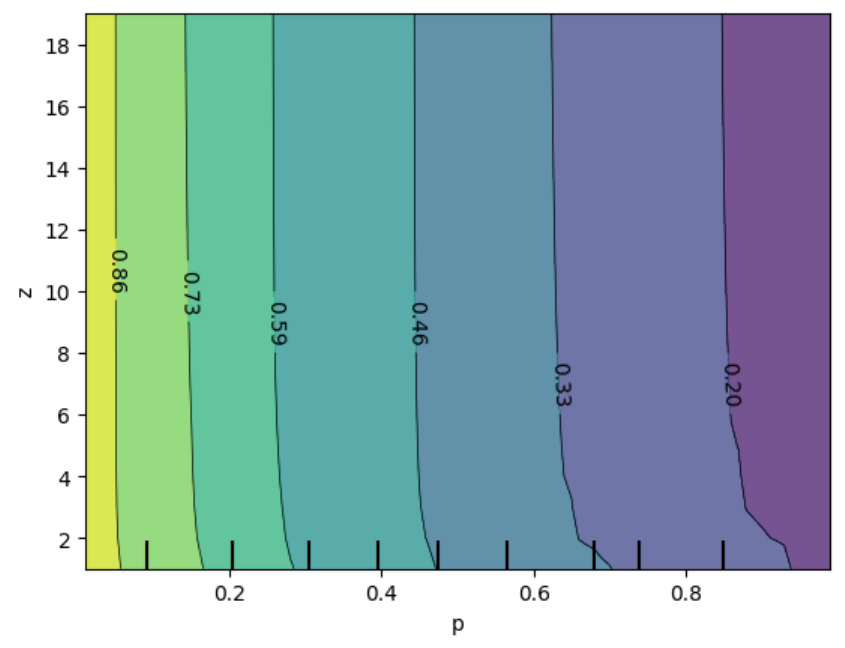}
        \caption{Two-Way Partial Dependence Plot between $p$ and $z$}
        \label{fig:pdp1-4}
    \end{subfigure}
    
    \vspace{0.2cm}
    \begin{subfigure}[t]{\columnwidth}
        \centering
        \includegraphics[width=0.8\columnwidth, height=4.5cm]{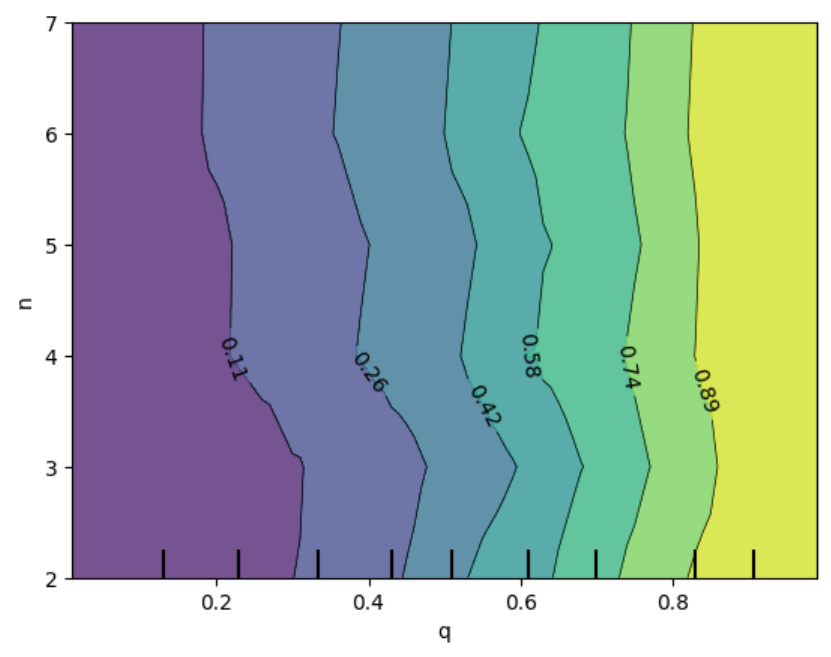}
        \caption{Two-Way Partial Dependence Plot between $q$ and $n$}
        \label{fig:pdp1-5}
    \end{subfigure}

\end{figure}

\begin{figure}[H]\ContinuedFloat
    \clearpage
    \begin{subfigure}[t]{\columnwidth}
        \centering
        \includegraphics[width=0.8\columnwidth, height=4.5cm]{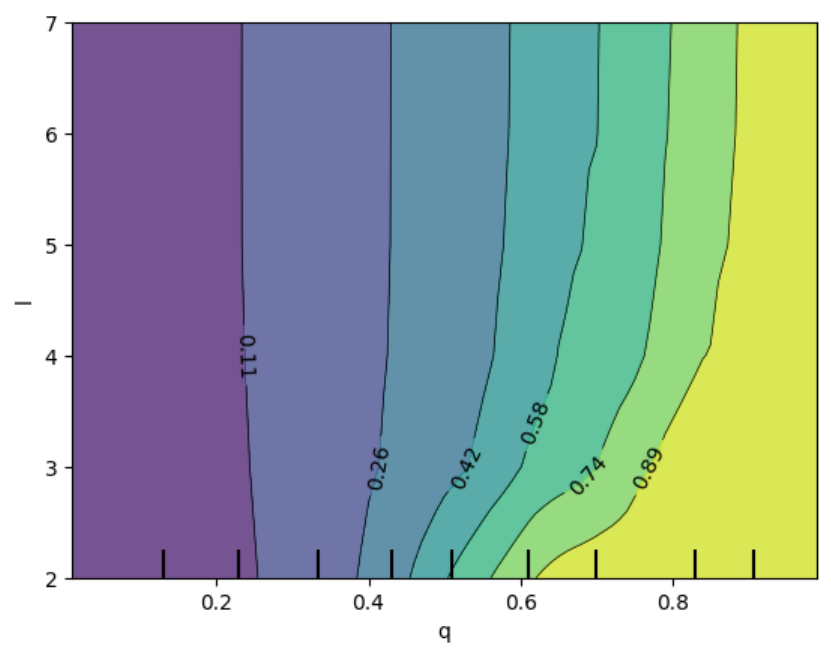}
        \caption{Two-Way Partial Dependence Plot between $q$ and $l$}
        \label{fig:pdp2-1}
    \end{subfigure}
    
    \vspace{0.2cm}
    \begin{subfigure}[t]{\columnwidth}
        \centering
        \includegraphics[width=0.8\columnwidth, height=4.5cm]{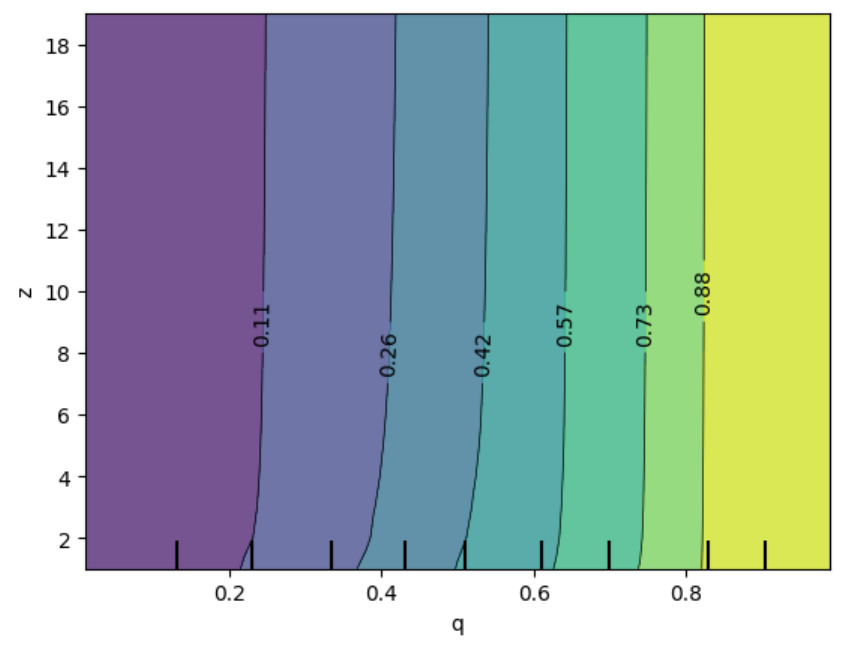}
        \caption{Two-Way Partial Dependence Plot between $q$ and $z$}
        \label{fig:pdp2-2}
    \end{subfigure}
    
    \vspace{0.2cm}
    \begin{subfigure}[t]{\columnwidth}
        \centering
        \includegraphics[width=0.8\columnwidth, height=4.5cm]{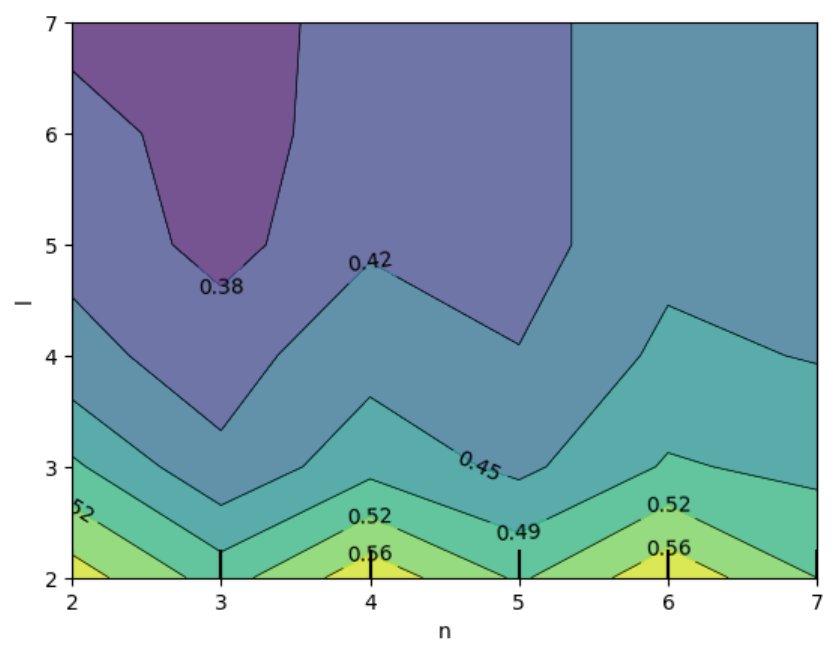}
        \caption{Two-Way Partial Dependence Plot between $n$ and $l$}
        \label{fig:pdp2-3}
    \end{subfigure}
    
    \vspace{0.2cm}
    \begin{subfigure}[t]{\columnwidth}
        \centering
        \includegraphics[width=0.8\columnwidth, height=4.5cm]{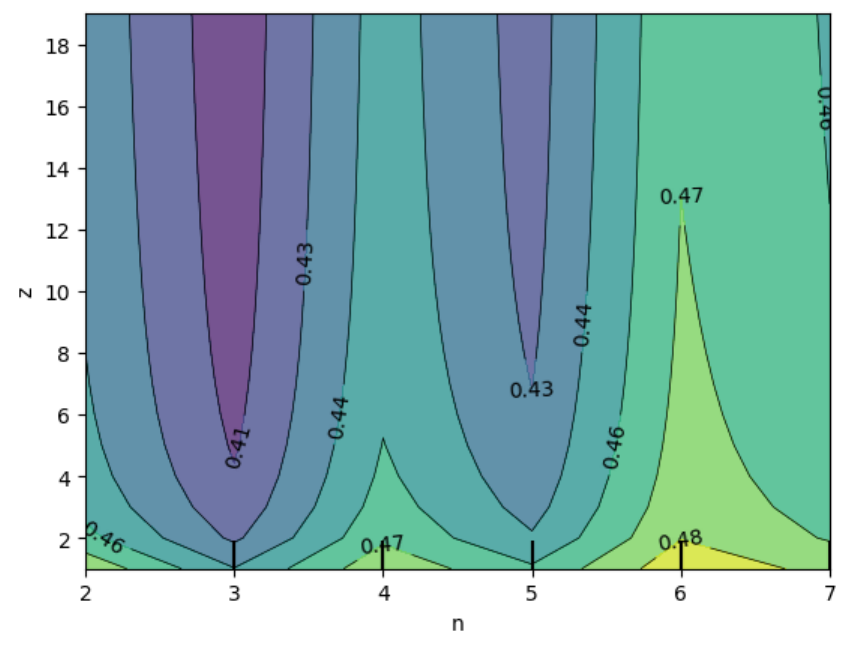}
        \caption{Two-Way Partial Dependence Plot between $n$ and $z$}
        \label{fig:pdp2-4}
    \end{subfigure}
    
    \vspace{0.2cm}
    \begin{subfigure}[t]{\columnwidth}
        \centering
        \includegraphics[width=0.8\columnwidth, height=4.5cm]{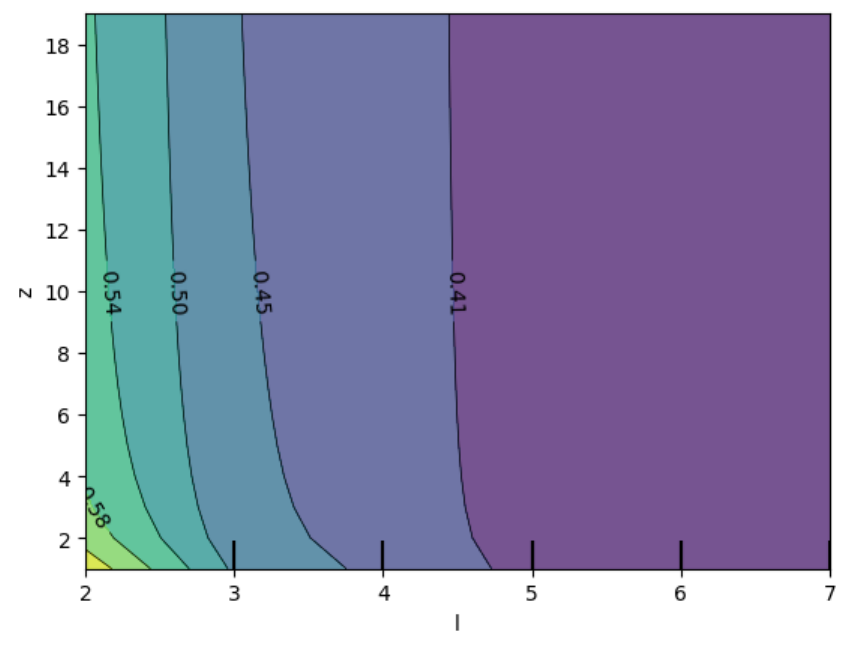}
        \caption{Two-Way Partial Dependence Plot between $l$ and $z$}
        \label{fig:pdp2-5}
    \end{subfigure}
    \caption{Two-Way Interaction Partial Dependence Plots.}
    \label{fig:pdp2}
\end{figure}

\subsection*{Asymptotic Analysis}

The asymptotic behavior of the probability $P(e|\lambda)$ at the extreme values of the parameters provides insights into how the system performs in limiting cases. These extreme values reveal the stability or vulnerability of the model in ideal or worst-case scenarios. The table below summarizes the limiting behavior of $P(e|\lambda)$ for different parameters:

\begin{table}[H]
\centering
\begin{tabular}{|C{1cm}|C{6cm}|}
\hline
\multicolumn{1}{|c|}{Extreme Values} & \multicolumn{1}{c|}{Limiting Value of $P(e|\lambda)$} \\
\hline
$p \rightarrow 0$ & $1.0$ \\ \hline
$p \rightarrow 1$ & \begin{itemize}
    \item  Case I: $(1-q)<q^d$
    \begin{itemize}
        \item[o] $P(e|\lambda) = 1.0$
    \end{itemize}
    \item  Case II: $n$ is even and $(1-q)^2 \ge q^d$
    \begin{itemize}
        \item[o] $\bar q_z = \left(\frac{q}{1-q^2}\right)^d$
        \item[o] $\lambda = z\bar q_z$
        \item[o] $P(e|\lambda) = 1 -  \sum\limits_{k=0}^{z}\frac{\lambda^k e^{-\lambda}}{k!}(1 - \bar{q}_{z-k})$
    \end{itemize}
    \item  Case III: $n$ is odd and $(1-q)^2 \ge q^d$
    \begin{itemize}
        \item[o] $\bar q_z = \left(\frac{q}{1-q}\right)^d$
        \item[o] $\lambda = z\bar q_z$
        \item[o] $P(e|\lambda) = 1 -  \sum\limits_{k=0}^{z}\frac{\lambda^k e^{-\lambda}}{k!}(1 - \bar{q}_{z-k})$
        \item[ ]
    \end{itemize}
\end{itemize} \\
\hline
$q \rightarrow 0$ & $0.0$ \\ \hline
$q \rightarrow 1$ & $1.0$ \\
\hline
$n \rightarrow 1$ & \begin{itemize}
    \item Case I: $q > \frac{1}{2}$
    \begin{itemize}
        \item[o] $P(e|\lambda) = 1.0$
    \end{itemize}
    \item Case II: $q \le \frac{1}{2}$
    \begin{itemize}
        \item[o] $\lambda = z\frac{q}{1-q}$
        \item[o] $P(e|\lambda) = 1 -  \sum\limits_{k=0}^{z}\frac{\lambda^k e^{-\lambda}}{k!}(1 - \frac{q}{1-q})$
        \item[ ] 
    \end{itemize}
\end{itemize} \\ \hline
$n \rightarrow \infty$ & $1.0$ \\
\hline
$z \rightarrow 0$ & $1.0$ \\ \hline
$z \rightarrow \infty$ & $0.0$ \\
\hline
$l \rightarrow 1$ & \begin{itemize}
    \item Case I: $q > \frac{1}{2}$
    \begin{itemize}
        \item[o] $P(e|\lambda) = 1.0$
    \end{itemize}
    \item Case II: $q \le \frac{1}{2}$
    \begin{itemize}
        \item[o] $\lambda = z\frac{q}{1-q}$
        \item[o] $P(e|\lambda) = 1 -  \sum\limits_{k=0}^{z}\frac{\lambda^k e^{-\lambda}}{k!}(1 - \frac{q}{1-q})$
        \item[ ]
    \end{itemize}
\end{itemize} \\ \hline
$l \rightarrow \infty$ & $1.0$ \\
\hline
\end{tabular}
\caption{Limiting values of $P(e|\lambda)$ at extreme values of attributes}
\label{table:probs}
\end{table}

The asymptotic behavior reveals several key insights:\\

1. \textbf{Effect of $p$ (Probability of Correct Verification)}:
   - As $p \rightarrow 0$, meaning no honest verifications occur, the probability of adversary success approaches 1, as expected.
   - As $p \rightarrow 1$, meaning perfect honest verifications occur, the probability of adversary success drops to 0, indicating a perfectly secure system.\\

2. \textbf{Effect of $q$ (Probability of Incorrect Verification)}:
   - As $q \rightarrow 0$, meaning there are no incorrect verifications, the adversary cannot succeed ($P(e|\lambda) = 0$).
   - As $q \rightarrow 1$, where all verifications are incorrect, the adversary always succeeds ($P(e|\lambda) = 1$).\\

3. \textbf{Effect of $n$ (Number of Verifiers)}:
   - For $n \rightarrow 1$, if $q > \frac{1}{2}$, the majority of verifiers are incorrect, leading to adversary success ($P(e|\lambda) = 1$).
   - As $n \rightarrow \infty$, even with a mixed group of verifiers ($p, q \in (0,1)$), the probability of adversary success remains high ($P(e|\lambda) = 1$) unless verification is near perfect.\\

4. \textbf{Effect of $z$ (Number of Honest Verifications)}:
   - As $z \rightarrow 0$, with no honest verifications, the adversary always succeeds ($P(e|\lambda) = 1$).
   - As $z \rightarrow \infty$, with many honest verifications, the adversary's success probability drops to zero ($P(e|\lambda) = 0$).\\

5. \textbf{Effect of $l$ (Height of the Hierarchy)}:
   - For $l \rightarrow 1$, if the majority of verifications are incorrect ($q > \frac{1}{2}$), the adversary succeeds.
   - As $l \rightarrow \infty$, with infinite levels of hierarchy but flawed verification ($p, q \in (0,1)$), the adversary still succeeds.

Overall, the limiting behavior suggests that the system becomes more secure with higher honest participation ($p \rightarrow 1$, $z \rightarrow \infty$) but becomes vulnerable in cases of widespread incorrect verification ($q \rightarrow 1$, $n \rightarrow \infty$, $l \rightarrow \infty$). In addition to these cases, there are some cases where the value of $P(e | \lambda)$ is specified as an expression. In those cases, the exact value will depend on the other parameters.

%% file: Sections/07.conclusion.tex
This research highlights the importance of leveraging hierarchical structures in Distributed Digital Ledgers (DDD) to strengthen multilevel verification processes. By aligning these natural hierarchies within domains like banking and distributed applications, the proposed approach significantly enhances security, scalability, and system resilience. Through detailed asymptotic analysis, we demonstrate that as the number of honest verifications increases, the probability of adversary success approaches zero, ensuring robust fraud prevention. The framework’s adaptability also extends to Centralized Digital Distributed (CDD) ledgers, which offer comparable security benefits within smaller, more controlled networks, making them ideal for scenarios where full decentralization is impractical.\\

Our approach underscores the versatility of both DDD and CDD ledgers, providing a scalable, adaptable solution to secure hierarchical multilevel verification in complex, dynamic systems. The combination of decentralized structure with multilevel security protocols positions this framework as a reliable and efficient solution across various industries.\\

In future work, we aim to extend this model by introducing hierarchies that evolve as a function of time. Adding temporal elements to the hierarchical verification process will enhance the system’s ability to adapt dynamically to changing conditions, further improving security and efficiency.